# Equivalence of Classical Statistics and Quantum Dynamics For Bosonic Field Theories

Paul J. Werbos[1] and Ludmilla Dolmatova Werbos[2]
[1]National Science Foundation[*], Room 675, Arlington, VA 22230
[2]IntControl, Arlington, VA 22203

## ABSTRACT

Years ago, Einstein asserted that it is too soon to give up on modeling the universe as a system of continuous force fields governed by partial differential equations (PDE). He argued that the complex laws of quantum dynamics might be derived someday as the laws of evolution of the *statistics* of PDE.

Most physicists gave up this hope years ago. They often cite a famous book by J.S. Bell. Bell argues that experiments based on a theorem of Clauser et al (often called "Bell's Theorem") prove that there cannot exist any kind of objective reality at all, or else that the laws of evolution must be nonlocal, leading to elaborate speculations about parallel universes and so on.[1]. Physicists of the Einstein school were stymied by the problem of "closure of turbulence" or "infinite regress" which makes it difficult to compute the statistics predicted by PDE without introducing extraneous apriori assumptions. The usual stoichiometric equations of thermodynamics, which were the foundation of the early work of Prigogine, were originally derived from the approximation of point particles in a free space of near-zero particle density [2]. Prigogine and others encountered a stubborn problem of "infinite regress" from higher moments to lower moments when they tried to start out from the more general assumption of continuous fields and interactions.

A recent paper (published in the International Journal of Bifurcation and Chaos[3], archived at arXiv.org as quant-ph 0309031) suggested that Einstein may have been right after all. It showed how to encode the statistical moments ("correlations") of bosonic PDE or ODE systems into a new mathematical object, the classical density matrix $\rho$. It provided a tutorial on field operators, and extensive details for bosonic ODE, ODE derived from the Lagrangian:

$$L = \frac{1}{2}\sum_{j=1}^{n}(\dot{\varphi}_j \pi_j - \dot{\pi}_j \varphi_j) - H(\underline{\varphi}, \underline{\pi})$$

where H is the Hamiltonian and $\underline{\varphi}$ and $\underline{\pi} \in R^n$. It proved that the usual field operators $\underline{\Phi}$ and $\underline{\Pi}$ defined in the standard text of Weinberg, applied to the classical density matrix $\rho$ as in quantum theory (Tr($\rho\underline{\Phi}$)), yield the classical expected values of $\underline{\varphi}$ and $\underline{\pi}$, *and* obey the quantum dynamical law given by Weinberg. It explained how "Bell's Theorem" experiments and the like result from differences in assumptions about *measurement* rather than differences in dynamics, and discussed the PDE case. It provided an updated version of the Backwards Time Interpretation of quantum mechanics first published in 1973 [4]; according to the updated version[3,5], subtle interactions between time-symmetry at the microscopic level and time-forwards thermodynamic effects at the macroscopic level are essential to a complete understanding of quantum measurement.

---
[*] The views herein are those of the authors and do not represent their employers.

Nevertheless, discrepancies between classical statistics and quantum dynamics do exist in the general case. This paper derives the free space master equation governing ρ. It calculates the classical/quantum discrepancies in predicting the expected time flux <$\partial_t g(\boldsymbol{\varphi},\boldsymbol{\pi})$> for a *general* observable g($\boldsymbol{\varphi},\boldsymbol{\pi}$). It shows that we can easily reformulate the classical field theory (by rescaling the field components $\varphi_j$ and $\pi_j$ without changing the *content* of the system) so as to zero out the major discrepancy terms. Similar reformulations mirroring the usual renormalization procedures of quantum theory may zero out the small higher-order discrepancies *in the case where* the quantum version is renormalizable or "finite" (requirements for mathematical validity of a quantum field theory). Proper scaling of $\boldsymbol{\varphi}$ and $\boldsymbol{\pi}$ is analogous to proper scaling of data prior to statistical analysis. An alternative method given in the Appendix of this paper may provide a more general method for the closure of turbulence. This is an upgraded version of quant-ph 0309087.

**1. Background, Summary and Concepts**

Before describing the new results reported in the Abstract, we should review a few major points.

First, the classical theories under discussion here *are not* the same as the "classical limit as h→0" discussed in elementary textbooks. We are discussing continuous field theories in the spirit of Einstein, not Lorentzian point particles. As an example, the classical sine-Gordon equation implicitly contains "h" as a term in the classical PDE, even though we often write that equation in $\hbar=c=1$ units; this should be clear from the role of the Klein-Gordon operator in that equation. The quantum sine-Gordon equation also uses the *normal-form* Hamiltonian [6], and results in soliton masses exactly the same as in the classical version.

Second, Bell's Theorem essentially rules out classical assumptions about *measurement*. It does not rule out equivalence or near-equivalence between classical statistical dynamics and quantum dynamics. The relationship between quantum dynamics and quantum measurement in Bell's theorem experiments is discussed at length in [3,5]. The equivalence between classical and quantum dynamics is already understood to be an important tool for researchers seeking effective numerical tools to compute the predictions of quantum dynamics [7]. "Phase space techniques" or "quantum trajectory simulations" have become an essential tool for practical work in quantum optics.

Third, many mainstream physicists now agree that measurement rules should be *derived* as an emergent consequence of quantum dynamics and of limited other information such as simple boundary conditions "at infinity" or spontaneous symmetry breaking on a cosmological scale. For example, the "measurement" effects of a polarizer in quantum computing are predicted more accurately by representing the polarizer as a solid state object [8,9], and invoking metaphysical observer concepts only for times after the experiment. In the Von-Neumann/Wigner view of quantum mechanics, we do better and better by restricting the metaphysical "observer effects" to distant times (later than $t_+$ or earlier than $t_-$) and sending $t_+$ and $t_-$ to $\pm\infty$. If we adopt this viewpoint, then we would expect to derive the *same* measurement rules for the classical density matrix $\rho$ at finite times, if it obeyed the same dynamics as the quantum density matrix! From Bell's Theorem work, we know that these measurement rules must violate classical "common sense" assumptions about time-forwards causality (as does quantum theory itself [10,11]); however, because they would be the same as the usual quantum rules, they would become equally credible as a theory of physics, so far as Bell's Theorem is concerned. Other traditional paradoxes related to measurement effects, such as the Kochen-Specker paradox, follow the same logic.

Fourth, the equivalence conjectured here would be limited to finite (i.e. "renormalizable with finite renormalizations") bosonic field theories. However, well-known physicists like Wilczek discovered long ago that certain subtle properties of spin and statistics do allow fermionic theories to be equivalent to bosonic theories, under certain circumstances. A large but fragmented literature on bosonization has resulted [12]. One important school of research, building on the work of Witten [13], claims to have proven that mixed Fermi-Bose theories over ordinary Minkowski space can be represented equivalently as bosonic theories[14].If so, then the standard model of physics could itself be represented equivalently as a bosonic field theory. The standard model is not a finite theory, but there are good reasons to want to modify it to make it finite, even from the traditional viewpoints of quantum theory.

This paper will begin, in section 2, by deriving the dynamics of the classical density matrix ρ. To make the discussion easier to follow, we will focus on the case of field theories in 0+1 dimensions (ODE); it is straightforward but tedious to extend the results to the PDE case, using the procedures given in [3]. *There will be frequent reference to equations from* [3]; thus for example we will refer to equation 7 of [3] as "IJBC-7." The discussion here will actually be fairly self-contained, for those who are familiar with the kind of mathematical tools used in modern quantum optics [8]; however, see [3] for a self-contained review of the tools used here.

The classical density matrix ρ does not obey the usual Liouville equation of QFT in the general case. Instead, it obeys a "free space master equation." Master equations are now the standard way to describe the quantum dynamics of many-body systems [8,9]. They provide a way to describe the *dissipative* effects which occur, in forwards time, for objects embedded in a macroscopic world governed by a forwards arrow of time. The usual Liouville equation of QFT is not dissipative.

But how is it possible that a dissipative kind of equation emerges *in free space*, for the statistics of time-symmetric PDE? Naively, one might guess that there is a kind of spontaneous symmetry breaking at work. But actually, there are two effects at work here.

First, there is a kind of excess information effect. Crudely speaking, the asymmetry of the free space master equation is analogous to the time-asymmetry of the usual stoichiometric reaction equations which result from analyzing time-symmetric random collisions and reactions of molecules floating in low-density space [2]. Some people have imagined that these stoichiometric equations are an example of spontaneous symmetry breaking – but they are not. The asymmetry in the statistical equations results from *allowing for the possibility* that the initial time state might be anything at all, while insisting that the system goes towards equilibrium in infinite forwards time. In short, *the asymmetry is an artifact of assumed boundary conditions*. In a similar way, the free space master equations are correct but *too general*, insofar as they allow for initial time conditions far out of thermodynamic equilibrium.

Secondly, the free space master equations give the exact dynamics for the moments $\rho$ for any statistical ensemble of states of the classical system. For such an ensemble, $\rho$ will always be positive definite and Hermitian. For such matrices $\rho$, we will write $\rho \in PR$, where "PR" stands for "physically realizable." But *not all positive definite Hermitian matrices* actually represent legitimate matrices of moments. This is related to the problem of moments in statistics [15]. In effect, the free space master equations, as written here, appear to allow for matrices ρ which are not even physically realizable. Again, they appear too general.

In order to address both of these issues, we face an obvious question: would we get back to the usual quantum dynamics if we *threw out* the normally meaningless additional information?

Sections 3 and 4 of this paper will argue that the answer to this question is essentially "yes," subject to certain caveats. When we first considered this question, we hoped that a simpler analysis might be enough to resolve it. Imitating the general approach of Weinberg [16, equations 7.1.28 and 7.1.29] and of Glimm and Jaffe, we hoped to enforce the usual dynamics *by definition*. More precisely, we planned to define a homomorphic mapping from the classical density matrix to a new matrix matching the quantum density matrix. For any pure state $\{\varphi(\underline{x}),\pi(\underline{x})\}$, we might first define:

$$\rho_0(\Delta) = \tfrac{1}{\Delta} \int_0^\Delta e^{i(H_n - E)t} \rho e^{-i(H_n - E)t} \, dt \tag{1}$$

where E is the total energy of the state and $H_n$ is the normal form Hamiltonian [3]. For an ensemble of states, we may define $\rho_0(\Delta)$ as the expectation value of $\rho_0(\Delta)$ across the ensemble. And then we may define:

$$\rho_Q = \lim_{\Delta \to \infty} \frac{\rho_0(\Delta)}{\text{Tr}(\rho_0(\Delta))} \tag{2}$$

(Alternatively, the expectation could be taken over the $\rho_Q$ for pure states.) The result would be a matrix $\rho_Q$ which looks much more like a normal density matrix of QFT! *By construction*, there would be no discrepancy between the prediction of observables based on this $\rho_Q$ and predictions based on QFT. Probably this procedure would also result in a situation where *all* nonnegative Hermitian matrices of unit trace correspond to the $\rho_Q$ of an ensemble of states ($\underline{\varphi}$, $\underline{\pi}$). But the integration over time in equation (1) may not really reflect the dynamics of the original classical field theory. For reasons beyond the scope of this paper, this first approach basically turns out to be an indirect back door to the more direct approach to be taken here.

This paper will explore a more direct approach to classical/quantum equivalence.. As in [3], we will explore whether the entire physics can be built up in Heisenberg's manner, from the dynamics of the observables, which are *dual* to $\rho$. In earlier work [5], we found that Fock space vectors or matrices which are *dual* to vectors or matrices of moments do not have the same completeness problems. Thus in section 3, we will evaluate whether the predictions of the classical theory for

$$< \dot{g} > = < \partial_t \langle g(\{\underline{\varphi}(x), \underline{\pi}(x)\}) \rangle > \tag{3}$$

match the corresponding predictions of quantum field theory (QFT)

$$\hat{\dot{g}} = \partial_t < g(\rho) > = \partial_t (\mathrm{Tr}(\dot{\rho} g_n)) \tag{4}$$

for any observable analytic function $g(\underline{\varphi}, \underline{\pi})$, where $g_n$ is the normal form field operator corresponding to that observable and where the dynamics of $\rho$ in equation 4 are calculated from the usual Liouville equation of QFT, starting from a classical density matrix $\rho \in PR$. Section 3 will calculate the difference between these two estimates of $\partial_t g$, for the ODE case (which embodies all the essential difficulties), for the case of field theories where the terms are third order or lower polynomials (like QED).

It turns out that these two predictions *do not exactly agree*, in the general case. In [3], it was proved that they do agree for the simple choices $g=\varphi_j$ or $g=\pi_k$, which are the usual starting points for discussion of the Heisenberg picture[16]. The conventional procedure of using the field operator g(**Φ**, **Π**) itself instead of $g_n$(**Φ**, **Π**) will not be explored here, because that g(**Φ**, **Π**) is not well-defined in the general case (due to commutativity issues), and because my calculations so far show that it has no advantages over $g_n$ even in cases where it is well-defined.

Section 4 will explore a different approach to explaining the discrepancies found in section 3. To begin with, note that QFT basically builds up all of its predictions by bootstrapping its ability to predict two things: (1) bound states (spectra); and (2) scattering – S matrices. In fact, all of the physics is contained in the S matrix for elementary particles, since bound states all appear as intermediate states in some kind of scattering process. Thus equivalent prediction of such S matrices and scattering states would imply equivalent quantum dynamics.

QFT has seen a wide variety of ways to define and calculate S matrices. But even today, the bulk of the successful, twelve-digits-precision predictions of QFT come from the predictions of quantum electrodynamics (QED), which fall out from the oldest S-matrix formalisms [17]. The most coherent description of what the S matrix actually means, so far as we know, comes from Dyson's paper [18], which refers to Heisenberg's original notion of stationary scattering states. Scattering states are simply *equilibrium states*, similar to bound states but not required to be localized in space; the boundary conditions are extended, to allow for a steady influx and outflux of particles.

Thus section 4 will explore the following notion of equivalence. Suppose that ρ represents an irreducible equilibrium ensemble (IEE) of the master equation; in other words, suppose that ρ represents an ensemble of pure states, all of which possess the same energy E (and identical values of certain other conserved quantities, depending on the field theory). Section 4 will start out by asking whether the discrepancies calculated in section 3 could be zero in the general case, when we exploit information about being in equilibrium.

The answer here is again negative, for the general case. However, consideration of a simple harmonic oscillator system brings out a fascinating result: there is an exact equivalence *only if* the classical representation of the field is scaled to make the "physical mass" equal the "bare mass"! In other words, such a field scaling or classical renormalization zeroes out the leading (second-order) term in the expression for the discrepancy given in section 3. Furthermore, the leading third-order discrepancy term directly represents "triple creation" or "triple annihilation" events in the Hamiltonian, events which are *thrown out* in actual calculations in QED (as are the traditional zero point energy terms [3]). There is still an infinite series of higher-order discrepancy terms, which appear similar in character to the infinite series of Feynman diagrams in S matrix calculations; thus it is reasonable to conjecture that the discrepancies may *all* disappear (or become very small) after we scale the classical fields in a way which reflects the full set of renormalizations used in QFT. Because quantum field theories which are not finite or renormalizable are arguably meaningless, we should not be surprised if exact equivalence to a meaningful classical theory only occurs when the quantum theory itself is finite or renormalizable.

Section 4 has assumed that calculations in QFT are based on the normal-form Hamiltonian, $H_n$, as is the case for practical calculations with Feymann diagrams in quantum electrodynamics. This assumption was discussed in [3], but has broad implications beyond the scope of this paper or of anything else written to date.

The Appendix will briefly describe an alternative approach to quantum-classical equivalence which appear promising but which we have not had time to fully explore.

**2. Derivation of the Dynamics (Master Equation) For ρ in the ODE Case**

**2.1. Review of Some Definitions and Results from IJBC**

Let us consider the 0+1 dimensional "field theory" defined by equation IJBC-7:

$$L = \tfrac{1}{2}\sum_{j=1}^{n}(\dot{\varphi}_j \pi_j - \dot{\pi}_j \varphi_j) - H(\underline{\varphi},\underline{\pi}) \tag{5}$$

A pure state of this system at any time t is specified by specifying $\underline{\varphi}(t)$ and $\underline{\pi}(t)$. We may define a kind of pseudo-wave-function $\underline{w}$ for a pure state of this system by IJBC-75:

$$\underline{w}(\underline{\varphi},\underline{\pi}) = \exp\left(\frac{1}{\sqrt{2}}\sum_{j=1}^{n}\left((\varphi_j + i\pi_j)a_j^+ - \tfrac{1}{2\sqrt{2}}(\varphi_j^2 + \pi_j^2)\right)\right)|0\rangle \tag{6}$$

(Note that there was a minor typo in this equation as printed in IJBC, corrected in the version at arXiv.org.) The classical density matrix ρ for an ensemble of pure states is defined in IJBC-74 as:

$$\rho = \iint \underline{w}(\underline{\varphi},\underline{\pi})\underline{w}^H(\underline{\varphi},\underline{\pi})\Pr(\underline{\varphi},\underline{\pi})d^n\underline{\varphi}\,d^n\underline{\pi} \quad , \tag{7}$$

where the superscript H denotes the Hermitian conjugate (complex conjugate transpose). In this paper, we will write "ρ∈PR" to indicate matrices ρ which are in the set of physically realizable classical density matrices – matrices which represent a statistical ensemble of pure states, as in equation 7. Weinberg's field operators reduce in this case to IJBC-7 and 48:

$$\Phi_j = \frac{1}{\sqrt{2}}(a_j + a_j^+) \tag{8}$$

$$\Pi_j = \frac{1}{i\sqrt{2}}(a_j - a_j^+) \qquad (9)$$

IJBC proved a general result (IJBC-81) which trivially implies:

$$\text{Tr}(\rho H_n(\underline{\Phi},\underline{\Pi})) = <H(\underline{\varphi},\underline{\pi})> \qquad (10)$$

where $H_n$ is the result of substituting the field operators $\underline{\Phi}$ and $\underline{\Pi}$ in for $\underline{\varphi}$ and $\underline{\pi}$ in $H(\underline{\varphi},\underline{\pi})$ *and of* replacing ordinary multiplication in the classical expression H by the *normal product*. $H_n$ is the normal form Hamiltonian. Likewise, the argument of section 4.2 of IJBC tells us that $H_n$ for any analytic function H may be written as:

$$H_n = \sum_\alpha C_\alpha H_{\alpha L} H_{\alpha R} = \sum_\alpha H_\alpha \qquad (11)$$

where "L" stands for "left" and "R" for "right"; where $\alpha$ is just an integer used to index the terms in the sum; where the sum is usually over a finite number of terms, but could be an infinite series; where $C_\alpha$ is any complex number; where $H_{\alpha L}$ is a product of creation operators; where $H_{\alpha R}$ is a product of annihilation operators; where each "$\alpha$" term in the sum can be matched with one and only one term in this series equal to its Hermitian conjugate; and where the expression $H_\alpha$ has been introduced for convenience in calculations.

## 2.2. The Free Space Master Equation

Let us define:

$$\rho' = i\sum_{j,\alpha} C_\alpha \left( a_j [a_j^+, H_{\alpha R}] \rho H_{\alpha L} - a_j^+ H_{\alpha R} \rho [a_j, H_{\alpha L}] \right) \qquad (12)$$

Then the free space master equation is:

$$\partial_t \rho = \rho' + (\rho')^H, \tag{13}$$

where the superscript H refers to the Hermitian conjugate. The remainder of this section will be devoted to proving that equation 9 is correct, for the systems described in section 2.1.

Master equations *in general* are equations which can be written as:

$$\partial_t \rho = F(\rho), \tag{14}$$

where ρ is some kind of density matrix and F is a linear transformation. Thus if wave functions are seen as a vector in Fock-Hilbert space, and density matrices as Hermitian nonnegative matrices over Fock-Hilbert space, functions F actually correspond to fourth-order tensors over Fock-Hilbert space. Equations 12 and 13, together, do constitute a master equation, by this definition; however, it is more convenient to write them in ordinary matrix/operator notation, in this paper.

**2.3. A Preliminary General Result for Pure States**

This section will build up to a new general result stated at the end of the section.

Let us start from the following result (IJBC-76) proved in section 4.2 of IJBC:

$$a_j \underline{w}(\underline{\varphi}, \underline{\pi}) = \left( \frac{\varphi_j + i\pi_j}{\sqrt{2}} \right) \underline{w}(\underline{\varphi}, \underline{\pi}) \tag{15}$$

Readers familiar with the methods used in [8] may see that equation 15 actually follows "by inspection" from equation 6.

For a pure state ($\underline{\varphi}, \underline{\pi}$), the definition in equation 7 reduces to:

$$\rho(\underline{\varphi},\underline{\pi}) = \underline{w}(\underline{\varphi},\underline{\pi})\underline{w}^H(\underline{\varphi},\underline{\pi}) \tag{16}$$

Multiplying equation 15 in the right by $\underline{w}^H$, and using equation 16, we deduce:

$$a_j \rho(\underline{\varphi},\underline{\pi}) = \left(\frac{\varphi_j + i\pi_j}{\sqrt{2}}\right)\rho(\underline{\varphi},\underline{\pi}) \tag{17}$$

The Hermitian conjugate of equation 17 is:

$$\rho(\underline{\varphi},\underline{\pi})a_j^+ = \left(\frac{\varphi_j - i\pi_j}{\sqrt{2}}\right)\rho(\underline{\varphi},\underline{\pi}) \tag{18}$$

To simplify the appearance of the equations, we may define some intermediate quantities (as in IJBC-91 and 92):

$$z_j = \tfrac{1}{\sqrt{2}}(\varphi_j + i\pi_j) \tag{19}$$

$$y_j = \tfrac{1}{\sqrt{2}}(\varphi_j - i\pi_j) \tag{20}$$

Because $z_j$ is just a scalar (for a pure state), equations 17 and 19 tell us that:

$$a_j^2 \rho(\underline{\varphi},\underline{\pi}) = a_j(a_j \rho(\underline{\varphi},\underline{\pi})) = a_j(z_j \rho(\underline{\varphi},\underline{\pi})) = z_j a_j \rho(\underline{\varphi},\underline{\pi}) = z_j^2 \rho(\underline{\varphi},\underline{\pi}) \tag{21}$$

More generally, by mathematical induction:

$$a_j^n \rho(\underline{\varphi},\underline{\pi}) = a_j(a_j^{n-1}\rho(\underline{\varphi},\underline{\pi})) = a_j(z_j^{n-1}\rho(\underline{\varphi},\underline{\pi})) = z_j^n \rho(\underline{\varphi},\underline{\pi}) \tag{22}$$

Likewise, for any four positive integers j, k, $i_j$ and $i_k$:

$$a_j^{i_j}(a_k^{i_k}\rho(\underline{\varphi},\underline{\pi})) = a_j^{i_j} z_k^{i_k} \rho(\underline{\varphi},\underline{\pi}) = z_j^{i_j} z_k^{i_k} \rho(\underline{\varphi},\underline{\pi}) \tag{23}$$

More generally, for any choice of complex number C and of integers in this equation:

$$C a_1^{i_1} a_2^{i_2} \ldots a_n^{i_n} \rho(\underline{\varphi},\underline{\pi}) = C z_1^{i_1} z_2^{i_2} \ldots z_n^{i_n} \rho(\underline{\varphi},\underline{\pi}) \tag{24}$$

As discussed in section 4.2 of IJBC, any polynomial or analytic function $f_R(\underline{a})$ may be represented as:

$$f_R(\underline{a}) = \sum_\alpha C_\alpha a_1^{i_{1\alpha}} a_2^{i_{2\alpha}} ... a_n^{i_{n\alpha}} \rho(\underline{\varphi},\underline{\pi}) \tag{25}$$

Combining equations 24 and 25, we may deduce, for any analytic function $f_R$:

$$f_R(\underline{a})\rho(\underline{\varphi},\underline{\pi}) = f_R(\underline{z})\rho(\underline{\varphi},\underline{\pi}) \tag{26}$$

From equations 19 and 20, the complex conjugate of $z_j$ is just $y_j$. Thus the Hermitian conjugate of equation 24 is:

$$\rho(\underline{\varphi},\underline{\pi})\overline{C}(a_1^+)^{i_1}(a_2^+)^{i_2}...(a_n^+)^{i_n} = \overline{C} y_1^{i_1} y_2^{i_2} ... y_n^{i_n} \rho(\underline{\varphi},\underline{\pi}) \tag{27}$$

where C-bar can be any complex number. Thus for any analytic function $f_L(\underline{a}^+)$, a representation like equation 25 may be combined with equation 27 to deduce:

$$\rho(\underline{\varphi},\underline{\pi}) f_L(\underline{a}^+) = f_L(\underline{y})\rho(\underline{\varphi},\underline{\pi}) \tag{28}$$

Furthermore, for any pair of analytic functions $f_L$ and $f_R$, we may deduce:

$$f_R(\underline{a})\rho(\underline{\varphi},\underline{\pi})f_L(\underline{a}^+) = f_R(\underline{a})(f_L(\underline{y})\rho(\underline{\varphi},\underline{\pi})) = f_L(\underline{y})f_R(\underline{z})\rho(\underline{\varphi},\underline{\pi}) \tag{29}$$

From section 4.2 of IJBC, we know that any analytic function $g(\underline{\varphi}, \underline{\pi})$ can be represented as:

$$g(\underline{\varphi},\underline{\pi}) = \sum_\alpha C_\alpha g_{\alpha L}(\underline{z}) g_{\alpha R}(\underline{y}), \tag{30}$$

for which the corresponding normal form field operator is:

$$g_n(\underline{\Phi},\underline{\Pi}) = \sum_\alpha C_\alpha g_{\alpha L}(\underline{a}^+) g_{\alpha R}(\underline{a}) \tag{31}$$

(Strictly speaking, we have not yet proven that this $g_n$ is the same as the $g_n$ defined in IJBC, but this is not hard to do, and is not strictly needed here). We will call equations 30 and 31 a *normal form decomposition* of $g$ and $g_n$ respectively. Combining equations 29 through 31, we immediately derive the main result of this subsection: for any analytic function $g(\underline{\varphi},\underline{\pi})$ and any pure state $(\underline{\varphi},\underline{\pi})$:

$$g(\underline{\varphi},\underline{\pi})\rho(\underline{\varphi},\underline{\pi}) = \sum_{\alpha} C_{\alpha} g_{\alpha R}(\underline{a})\rho(\underline{\varphi},\underline{\pi})g_{\alpha L}(\underline{a}^{+}) \tag{32}$$

The curious or skeptical reader may note that equation 10 and equation IJBC-81 can be derived simply by taking the trace of equation 32, using wall-known trace identities to prove the result for pure states, and taking expectation values to prove IJBC-81 for the general case of any statistical ensemble of states.

### 2.4. A Lemma For Dynamics of Pure States

For all pure states ($\underline{\varphi}$, $\underline{\pi}$) governed by the dynamics implied by section 2.1, we claim that:

$$\dot{\varphi}_j \rho = -\tfrac{i}{\sqrt{2}} \sum_{\alpha} C_{\alpha}\left([a_j^+, H_{\alpha R}]\rho H_{\alpha L} + H_{\alpha R}\rho[a_j, H_{\alpha L}]\right) \tag{33}$$

and:

$$\dot{\pi}_j \rho = -\tfrac{1}{\sqrt{2}} \sum_{\alpha} C_{\alpha}\left(-[a_j^+, H_{\alpha R}]\rho H_{\alpha L} + H_{\alpha R}\rho[a_j, H_{\alpha L}]\right) \tag{34}$$

Proof:

From Hamiltonian field theory, it is well known that:

$$\dot{\varphi}_j = \frac{\partial H}{\partial \pi_j} \tag{35}$$

$$\dot{\pi}_j = -\frac{\partial H}{\partial \varphi_j} \tag{36}$$

The previous paper also showed (IJBC-56 and 57, choosing "g" to be "H"):

$$[\Phi_j, H_n(\underline{\Phi},\underline{\Pi})] = i\left(\frac{\partial H}{\partial \pi_j}\right)_n \tag{37}$$

and:

$$[\Pi_j, H_n(\underline{\Phi},\underline{\Pi})] = -i\left(\frac{\partial H}{\partial \varphi_j}\right)_n \qquad (38)$$

where the subscript "n" again denotes the normal form operator for the function given in parentheses.

If we substitute equations 8 and 11 into the left-hand side of equation 37, we may deduce:

$$i\left(\frac{\partial H}{\partial \pi_j}\right)_n = \left[\tfrac{1}{\sqrt{2}}(a_j + a_j^+), \sum_\alpha C_\alpha H_{\alpha L} H_{\alpha R}\right] \qquad (39)$$
$$= \tfrac{1}{\sqrt{2}} \sum_\alpha C_\alpha \left([a_j, H_{\alpha L}] H_{\alpha R} + H_{\alpha L}[a_j^+, H_{\alpha R}]\right)$$

With minor manipulation, this becomes:

$$\left(\frac{\partial H}{\partial \pi_j}\right)_n = \tfrac{-i}{\sqrt{2}} \sum_\alpha C_\alpha \left([a_j, H_{\alpha L}] H_{\alpha R}\right) - \tfrac{i}{\sqrt{2}} \sum_\alpha C_\alpha \left(H_{\alpha L}[a_j^+, H_{\alpha R}]\right) \qquad (40)$$

Note that equation 40 already constitutes a normal form decomposition! More precisely, each sum on the right hand side of equation 40 has the required form for a term in the sum shown in equation 31, and the sum of two sums of terms of the required form is still a sum of terms of the required form. Therefore we may use equation 32 to deduce:

$$\dot{\varphi}_j \rho = \left(\frac{\partial H}{\partial \pi_j}\right)_n \rho = \tfrac{-i}{\sqrt{2}} \sum_\alpha C_\alpha \left(H_{\alpha R} \rho [a_j, H_{\alpha L}]\right) - \tfrac{i}{\sqrt{2}} \sum_\alpha C_\alpha \left([a_j^+, H_{\alpha R}] \rho H_{\alpha L}\right) \qquad (41)$$

This is clearly equivalent to equation 33. We have proved the validity of equation 33. Next, if we substitute equations 9 and 11 into the left hand side of equation 38, we may deduce:

$$-i\left(\frac{\partial H}{\partial \varphi_j}\right)_n = \left[\tfrac{1}{i\sqrt{2}}(a_j - a_j^+), \sum_\alpha C_\alpha H_{\alpha L} H_{\alpha R}\right] \qquad (42)$$

$$= \tfrac{1}{i\sqrt{2}} \sum_\alpha C_\alpha \left([a_j, H_{\alpha L}]H_{\alpha R} - H_{\alpha L}[a_j^+, H_{\alpha R}]\right)$$

which implies:

$$-\left(\frac{\partial H}{\partial \varphi_j}\right)_n = -\tfrac{1}{\sqrt{2}} \sum_\alpha C_\alpha \left([a_j, H_{\alpha L}]H_{\alpha R} - H_{\alpha L}[a_j^+, H_{\alpha R}]\right) \qquad (43)$$

Once again using equation 32, we may deduce:

$$\dot{\pi}_j \rho = -\left(\frac{\partial H}{\partial \varphi_j}\right)\rho = -\tfrac{1}{\sqrt{2}} \sum_\alpha C_\alpha \left(H_{\alpha R}\rho[a_j, H_{\alpha L}] - [a_j^+, H_{\alpha R}]\rho H_{\alpha L}\right) \qquad (44)$$

Since equation 44 is equivalent to equation 34, we have now proved both equations 33 and 34. QED.

### 2.5. Proof of Free Space Master Equation for Pure States

In order to prove that equations 12 and 13 are correct, for all statistical ensembles of states of the system governed by equation 5, we need only prove that they are correct for pure states ($\underline{\varphi}, \underline{\pi}$). More precisely – once we prove that equations 12 and 13 are correct for all pure states, we can multiply both equations by $\Pr(\underline{\varphi}, \underline{\pi})$ and integrate over $\underline{\varphi}$ and $\underline{\pi}$, which immediately proves the validity of the equations for statistical ensembles.

In order to prove equations 12 and 13 for a pure state, we will basically just work out what $\partial_t \rho$ is for a pure state. We may begin by invoking the chain rule:

$$\frac{\partial}{\partial t}\rho(\underline{\varphi}, \underline{\pi}) = \sum_{j=1}^n \left(\frac{\partial \rho}{\partial \varphi_j} \cdot \dot{\varphi}_j + \frac{\partial \rho}{\partial \pi_j} \cdot \dot{\pi}_j\right) \qquad (45)$$

In order to work out the first term in this expression, we first apply the chain rule to equation 16:

$$\frac{\partial \rho}{\partial \varphi_j} = \left(\frac{\partial}{\partial \varphi_j} w(\underline{\varphi},\underline{\pi})\right) w^H(\underline{\varphi},\underline{\pi}) + w(\underline{\varphi},\underline{\pi})\left(\frac{\partial}{\partial \varphi_j} w(\underline{\varphi},\underline{\pi})\right)^H \quad (46)$$

We may simply differentiate equation 6 in order to calculate:

$$\frac{\partial}{\partial \varphi_j} w(\underline{\varphi},\underline{\pi}) = \frac{1}{\sqrt{2}}\left(a_j^+ - \frac{1}{\sqrt{2}}\varphi_j\right) w(\underline{\varphi},\underline{\pi}) \quad (47)$$

Note that it is legitimate to differentiate equation 6 as we would differentiate a conventional algebraic expression because all of the creation operators and scalars here commute with each other.

If we substitute equation 47 into equation 46, and use the definition in equation 16 to simplify the results, we derive:

$$\frac{\partial \rho}{\partial \varphi_j} = \frac{1}{\sqrt{2}}\left(a_j^+ - \frac{1}{\sqrt{2}}\varphi_j\right)\rho + \frac{1}{\sqrt{2}}\rho(a_j - \frac{1}{\sqrt{2}}\varphi_j) = \frac{1}{\sqrt{2}}\left(a_j^+ \rho + \rho a_j\right) - \varphi_j \rho \quad (48)$$

Furthermore, by adding together equations 17 and 18 we deduce (for any pure state ρ(φ,π)):

$$a_j \rho + \rho a_j^+ = \sqrt{2}\varphi_j \rho \quad (49)$$

If we divide equation 49 by sqrt(2) and substitute into equation 48, we derive:

$$\frac{\partial \rho}{\partial \varphi_j} = \frac{1}{\sqrt{2}}\left(a_j^+ \rho + \rho a_j\right) - \frac{1}{\sqrt{2}}\left(a_j \rho + \rho a_j^+\right) = \frac{1}{\sqrt{2}}\left(a_j^+ - a_j\right)\rho + \frac{1}{\sqrt{2}}\rho\left(a_j - a_j^+\right) \quad (50)$$

In similar fashion, we may differentiate equation 6 again to calculate:

$$\frac{\partial}{\partial \pi_j} w(\underline{\varphi},\underline{\pi}) = \frac{1}{\sqrt{2}}\left(ia_j^+ - \frac{1}{\sqrt{2}}\pi_j\right) w(\underline{\varphi},\underline{\pi}) \quad (51)$$

If we substitute equation 51 into the equation for (∂ρ/∂π_j) analogous to equation 46, we derive an equation analogous to equation 48:

$$\frac{\partial \rho}{\partial \pi_j} = \frac{1}{\sqrt{2}}\left(ia_j^+ - \frac{1}{\sqrt{2}}\pi_j\right)\rho + \frac{1}{\sqrt{2}}\rho(-ia_j - \frac{1}{\sqrt{2}}\pi_j) = \frac{1}{\sqrt{2}}(ia_j^+\rho - i\rho a_j) - \pi_j\rho \quad (52)$$

Subtracting equation 18 from equation 17, we derive for any pure state ρ:

$$a_j\rho - \rho a_j^+ = i\sqrt{2}\pi_j\rho \quad (53)$$

Multiplying equation 53 by i/sqrt(2), and substituting into the last term in equation 53, we deduce:

$$\frac{\partial \rho}{\partial \pi_j} = \frac{1}{\sqrt{2}}(ia_j^+\rho - i\rho a_j) + \frac{i}{\sqrt{2}}(a_j\rho - \rho a_j^+) = \frac{i}{\sqrt{2}}(a_j^+ + a_j)\rho - \frac{i}{\sqrt{2}}\rho(a_j + a_j^+) \quad (54)$$

We have finally worked out all the terms we need in order to evaluate equation 45. First we may substitute equations 50 and 54 into equation 45 and multiply by sqrt(2) in order to simplify the result a little:

$$\sqrt{2}\dot{\rho} = \sum_{j=1}^{n}\left((a_j^+ - a_j)(\dot{\varphi}_j\rho) + (\dot{\varphi}_j\rho)(a_j - a_j^+) + i(a_j^+ + a_j)(\dot{\pi}_j\rho) - i(\dot{\pi}_j\rho)(a_j + a_j^+)\right) \quad (55)$$

Next we may take advantage of the lemma proved in section 2.4. We may substitute equations 33 and 34 into equation 55 to derive:

$$\sqrt{2}\dot{\rho} = \frac{i}{\sqrt{2}}\sum_{j,\alpha}C_\alpha\left(-(a_j^+ - a_j)([a_j^+, H_{\alpha R}]\rho H_{\alpha L} + H_{\alpha R}\rho[a_j, H_{\alpha L}])\right.$$
$$-([a_j^+, H_{\alpha R}]\rho H_{\alpha L} + H_{\alpha R}\rho[a_j, H_{\alpha L}])(a_j - a_j^+)$$
$$-(a_j^+ + a_j)(-[a_j^+, H_{\alpha R}]\rho H_{\alpha L} + H_{\alpha R}\rho[a_j, H_{\alpha L}])$$
$$\left.+(-[a_j^+, H_{\alpha R}]\rho H_{\alpha L} + H_{\alpha R}\rho[a_j, H_{\alpha L}])(a_j^+ + a_j)\right) \quad (56)$$

Collecting similar terms in equation 56 and dividing by sqrt(2), we derive:

$$\dot{\rho} = \tfrac{i}{2}\sum_{j,\alpha}C_\alpha\left(2a_j[a_j^+, H_{\alpha R}]\rho H_{\alpha L} - 2a_j^+ H_{\alpha R}\rho[a_j, H_{\alpha L}] - 2[a_j^+, H_{\alpha R}]\rho H_{\alpha L}a_j + 2H_{\alpha R}\rho[a_j, H_{\alpha L}]a_j^+\right)$$
$$(57)$$

Using the definition in equation 12, this can be expressed as:

$$\dot{\rho} = i\sum_{j,\alpha} C_\alpha \left(a_j [a_j^+, H_{\alpha R}]\rho H_{\alpha L} - a_j^+ H_{\alpha R}\rho[a_j, H_{\alpha L}]\right) + i\sum_{j,\alpha} C_\alpha \left(H_{\alpha R}\rho[a_j, H_{\alpha L}]a_j^+ - [a_j^+, H_{\alpha R}]\rho H_{\alpha L} a_j\right)$$

$$= \rho' + i\sum_{j,\alpha} C_\alpha \left(H_{\alpha R}\rho[a_j, H_{\alpha L}]a_j^+ - [a_j^+, H_{\alpha R}]\rho H_{\alpha L} a_j\right)$$

(58)

At first glance, equation 58 does not appear identical to equations 12 and 13, since the individual terms in the final summation are *not* the Hermitian conjugates of the corresponding terms in equation 12. However, in the specification of equation 11, it was required that each term α is matched to one and only one term equal to its Hermitian conjugate. This was required, in order to ensure that $H_n$ as a whole be Hermitian. (This was discussed in more detail in IJBC. Note that some terms α may be Hermitian, while all others come in pairs of Hermitian conjugates.) In other words, for each value of the index variable α there is one and only one β such that:

$$C_\alpha H_{\alpha L} H_{\alpha R} = \left(C_\beta H_{\beta L} H_{\beta R}\right)^H$$

(59)

Since $H_{\alpha L}$ and $H_{\alpha R}$ are specified to be strings of creation and annihilation operators, respectively, and likewise for β, we may deduce:

$$C_\alpha = \overline{C}_\beta$$

(60)

$$H_{\alpha L} = H_{\beta R}^H$$

(61)

$$H_{\alpha R} = H_{\beta L}^H$$

(62)

Substituting this into equation 58, we derive:

$$\dot{\rho} - \rho' = i\sum_{j,\alpha} \overline{C}_\beta \left(H_{\beta L}^H \rho [a_j, H_{\beta R}^H] a_j^+ - [a_j^+, H_{\beta L}^H]\rho H_{\beta R}^H a_j\right)$$

(63)

Strictly speaking, each "β" in equation 63 is an abbreviation for β(α), the β which corresponds to a particular choice of α. However, as we cycle through each value of α once and only once, we also cycle through each value of β once and only once. Thus equation 63 remains valid if we simply replace the "α" under the summation sign with "β." But in that case, "β" is just a bound variable; we know from formal logic [19] that the expression remains valid if we replace it with "α."

Thus we may deduce from equation 63 that:

$$\dot{\rho} - \rho' = i \sum_{j,\alpha} \overline{C}_\alpha \left( H^H_{\alpha L} \rho [a_j, H^H_{\alpha R}] a_j^+ - [a_j^+, H^H_{\alpha L}] \rho H^H_{\alpha R} a_j \right) \tag{64}$$

Since the right hand side of equation 64 is in fact the Hermitian conjugate of ρ' as defined in equation 12, we have now proved equation 13 for all pure states. As noted at the start of this section, this immediately proves equations 12 and 13 for all statistical ensembles. Q.E.D.

Remark: Under equation 58, it is easy to see that $\partial_t(Tr(\rho))=0$ for *all* matrices ρ, regardless of whether ρ is physically realizable (i.e., represents an ensemble of pure states) or not. However, this is not true for $\partial_t(Tr(\rho H_n))$, so far as we can tell, even though $H_n$ certainly is conserved for all physically realistic states.

## 3. Discrepancies: Classical Versus Quantum ġ For The Same ρ∈PR

### 3.1. How Discrepancy Will Be Calculated

The free space master equations (14, 15 and 58) appear very different from the usual Liouville equation of QFT, which in my view [3] should be written as:

$$\dot{\rho} = -i[H_n, \rho] \tag{65}$$

However, this difference in appearance would not be relevant to the actual physics, *if* predictions of the dynamics of actual *observables* g($\boldsymbol{\varphi}$, $\boldsymbol{\pi}$) were the same in quantum physics as in classical physics.

This section will calculate the discrepancy, $\hat{\dot{g}} - \dot{g}$, between the classical and the quantum-mechanical predictions of the change in an observable g, *when* we use the same density matrix $\rho \in$ PR as the starting point for the two predictions. In both cases, we will use the same operator $g_n$ ($\boldsymbol{\Phi}$, $\boldsymbol{\Pi}$) to define the observable. The prediction $\dot{g}$ will be defined as the prediction based on the free space master equations. The prediction $\hat{\dot{g}}$ will be based on the Liouville equation (65). As in section 2, all calculations will be performed for pure states $\rho(\boldsymbol{\varphi}, \boldsymbol{\pi})$; because of linearity in $\rho$, the generalization to all $\rho \in$ PR is trivial.

Please note that this *may not turn out to be* the best way to define the classical/quantum discrepancy in the end. Here we are calculating the discrepancy between QFT *and the theory that* the Heisenberg operators should be viewed as operators *operating on* $\rho$. We are evaluating the discrepancy when the "same" Hamiltonian is used both on the classical side and on the quantum side. But there are alternative possible neoclassical theories of physics. For example, the Appendix suggests a kind of mapping from classical field theories to *different* but legitimate operator field theories. The analysis of such theories would be a worthwhile exercise, but it is beyond the scope of this paper.

In [3], it was proved that the discrepancy equals zero, *for the case* of $\varphi_j$ or $\pi_j$ as observables. However, the discrepancy is not always zero for all observables g($\boldsymbol{\varphi}$, $\boldsymbol{\pi}$).

In order to get a basic understanding of these discrepancies, we will first calculate them for the case where **φ** and **π** are simply scalars, $\varphi$ and $\pi$, and where $H_n$ is of limited order (like QED); more precisely, section 3.2 will calculate $\hat{\dot{g}} - \dot{g}$ for the case where n=1 and where:

$$[a,[a,[a,[a,H_n]]]] = [[[[H_n,a^+],a^+],a^+],a^+] = 0 \tag{66}$$

Section 3.3 will extend this slightly by considering the case where n=2, which is complicated enough to show how interactions between different components of **φ** and **π** affect the discrepancy.

## 3.2. Calculation of Discrepancy For n=1 Under Equation 66

Claim: For any pure state $\rho(\varphi, \pi)$ governed by the dynamics described in section 2.1, subject to equation 66, the discrepancy obeys:

$$8i(\hat{\dot{g}} - \dot{g}) = -4i\left(\frac{\partial^2 g}{\partial\varphi\partial\pi}\right)\left(\frac{\partial^2 H}{\partial\varphi^2} - \frac{\partial^2 H}{\partial\pi^2}\right) + 4i\left(\frac{\partial^2 g}{\partial\varphi^2} - \frac{\partial^2 g}{\partial\pi^2}\right)\left(\frac{\partial^2 H}{\partial\varphi\partial\pi}\right)$$
$$+ \tfrac{1}{3}\left(\frac{\partial^3 g}{\partial\varphi^3} - 3\frac{\partial^3 g}{\partial\varphi\partial\pi^2}\right)\left(3i\frac{\partial^3 H}{\partial\varphi^2\partial\pi} - i\frac{\partial^3 H}{\partial\pi^3}\right) + \tfrac{1}{3}\left(-3i\frac{\partial^3 g}{\partial\varphi^2\partial\pi} + i\frac{\partial^3 g}{\partial\pi^3}\right)\left(\frac{\partial^3 H}{\partial\varphi^3} - 3\frac{\partial^3 H}{\partial\varphi\partial\pi^2}\right)$$
$$\tag{67}$$

The proof occupies the remainder of this section.

### 3.2.1. A Preliminary Result

Lemma: For all $H_n$ which obey equation 66,

$$[a^n, H_n] = n[a, H_n]a^{n-1} + \frac{n(n-1)}{2}[a,[a,H_n]]a^{n-2} + \frac{n(n-1)(n-2)}{6}[a,[a,[a,H_n]]]a^{n-3}, \tag{68}$$

where the term $a^{n-k}$ is interpreted as zero when k>n.

Proof by mathematical induction:

The validity of 68 for the case n=1 is obvious by inspection. To prove that equation 68 is valid for any n, when it is valid for n-1, we may use well-known commutator relations (IJBC-26) to calculate:

$$[a^n, H_n] = [a, H_n]a^{n-1} + a[a^{n-1}, H_n] \tag{69}$$

By the induction hypothesis (the assumption that equation 68 is valid for the case n-1), equation 69 may be expanded to:

$$[a^n, H_n] = [a, H_n]a^{n-1} + a(n-1)[a, H_n]a^{n-2} + a\frac{(n-1)(n-2)}{2}[a,[a,[H_n]]a^{n-3}$$
$$+ a\frac{(n-1)(n-2)(n-3)}{6}[a,[a,[a, H_n]]]a^{n-4} \tag{70}$$

We may expand each of the three rightmost terms, using the definition of a commutator:

$$a(n-1)[a, H_n]a^{n-2} = (n-1)[a,[a, H_n]]a^{n-2} + (n-1)[a, H_n]a^{n-1} \tag{71}$$

$$a\frac{(n-1)(n-2)}{2}[a,[a, H_n]]a^{n-3} = \frac{(n-1)(n-2)}{2}[a,[a,[a, H_n]]]a^{n-3} + \frac{(n-1)(n-2)}{2}[a,[a, H_n]]a^{n-2} \tag{72}$$

$$a\frac{(n-1)(n-2)(n-3)}{6}[a,[a,[a, H_n]]]a^{n-4} = \frac{(n-1)(n-2)(n-3)}{6}[a,[a,[a, H_n]]]a^{n-3} \tag{73}$$

Note that in equation 73 we have exploited equation 66 to justify treating the term involving [a,[a,[a,[a,H$_n$]]]] as zero.

Inserting equations 71,72 and 73 into 70 and collecting like terms, we deduce:

$$[a^n, H_n] = n[a, H_n]a^{n-1} + \left((n-1) + \frac{(n-1)(n-2)}{2}\right)[a,[a, H_n]]a^{n-2}$$
$$+ \left((n-1)(n-2)\left(\frac{1}{2} + \frac{n-3}{6}\right)\right)[a,[a,[a, H_n]]]a^{n-3} \tag{74}$$

This in turn reduces to equation 68, what we are trying to prove. QED.

Remark: equation 68 may also be seen as the operator form of the second through fourth terms of the Taylor expansion of the function H!

Corollary 1: For all $H_n$ which obey equation 66,

$$[(a^+)^m, H_n] = m(a^+)^{m-1}[a^+, H_n] - \frac{m(m-1)}{2}(a^+)^{m-2}[a^+,[a^+, H_n]] \\ + \frac{m(m-1)(m-2)}{6}(a^+)^{m-3}[a^+,[a^+,[a^+, H_n]]] \quad (75)$$

This is simply the Hermitian conjugate of equation 68, with n changed to m.

Corollary 2: By exploiting equations 68 and 75, we may deduce:

$$[(a^+)^m a^n, H_n] = [(a^+)^m, H_n]a^n + (a^+)^m[a^n, H_n] = \\ n(a^+)^m[a, H_n]a^{n-1} + \frac{n(n-1)}{2}(a^+)^m[a,[a, H_n]]a^{n-2} + \frac{n(n-1)(n-2)}{6}(a^+)^m[a,[a,[a, H_n]]]a^{n-3} \\ + m(a^+)^{m-1}[a^+, H_n]a^n - \frac{m(m-1)}{2}(a^+)^{m-2}[a^+,[a^+, H_n]]a^n + \frac{m(m-1)(m-2)}{6}(a^+)^{m-3}[a^+,[a^+,[a^+, H_n]]]a^n$$

(76)

### 3.2.2. Calculation of Discrepancy for g(φ,π) of the Form $z^n y^m$

Let us now calculate the discrepancy $\hat{\dot{g}} - \dot{g}$ for functions g of the form

$$g(\varphi, \pi) = z^n y^m, \quad (77)$$

where z and y are defined as in equations 19 and 20. This implies that the corresponding normal form operator is:

$$g_n = (a^+)^m a^n \quad (78)$$

First let us calculate the classical value for $\dot{g}$:

$$\dot{g} = (nz^{n-1} y^m)\dot{z} + (mz^n y^{m-1})\dot{y} \quad (79)$$

From IJBC-95 and 96, we know that:

$$\dot{z} = -i\operatorname{Tr}(\rho[a, H_n]) \tag{80}$$

$$\dot{y} = -i\operatorname{Tr}(\rho[a^+, H_n]) \tag{81}$$

Substituting 80 and 81, we get:

$$\dot{g} = -i(nz^{n-1}y^m)\operatorname{Tr}(\rho[a, H_n]) - i(mz^n y^{m-1})\operatorname{Tr}(\rho[a^+, H_n]) \tag{82}$$

Next, to work out the QFT prediction, we start from equation 65 and calculate:

$$\dot{\hat{g}} = \frac{\partial}{\partial t}\operatorname{Tr}(\rho g_n) = \operatorname{Tr}(\dot{\rho} g_n) = -i\operatorname{Tr}((H_n\rho - \rho H_n)g_n) = -i\operatorname{Tr}(\rho g_n H_n - \rho H_n g) = -i\operatorname{Tr}(\rho[g_n, H_n]) \tag{83}$$

Note that we have just exploited the well-known trace identity:

$$\operatorname{Tr}(ABC) = \operatorname{Tr}(BCA) = \operatorname{Tr}(CAB) \tag{84}$$

To evaluate the right-hand side of equation 83, we may simply recall the value of $g_n$ as given in equation 78, and exploit equation 76, to get:

$$\dot{\hat{g}} = -i\operatorname{Tr}\left(\rho\left(n(a^+)^m[a, H_n]a^{n-1} + \frac{n(n-1)}{2}(a^+)^m[a,[a, H_n]]a^{n-2} + \frac{n(n-1)(n-2)}{6}(a^+)^m[a,[a,[a, H_n]]]a^{n-3}\right.\right.$$
$$\left.\left.+ m(a^+)^{m-1}[a^+, H_n]a^n - \frac{m(m-1)}{2}(a^+)^{m-2}[a^+,[a^+, H_n]]a^n + \frac{m(m-1)(m-2)}{6}(a^+)^{m-3}[a^+,[a^+,[a^+, H_n]]]a^n\right)\right) \tag{85}$$

Exploiting equations 29 and 84 to simplify equation 85, and multiplying by i, we get:

$$i\dot{\hat{g}} = nz^{n-1}y^m \operatorname{Tr}(\rho[a, H_n]) + \frac{n(n-1)}{2}z^{n-2}y^m \operatorname{Tr}(\rho[a,[a, H_n]])$$
$$+ \frac{n(n-1)(n-2)}{6}z^{n-3}y^m \operatorname{Tr}(\rho[a,[a,[a, H_n]]])$$
$$mz^n y^{m-1} \operatorname{Tr}(\rho[a^+, H_n]) - \frac{m(m-1)}{2}z^n y^{m-2} \operatorname{Tr}(\rho[a^+,[a^+, H_n]])$$
$$+ \frac{m(m-1)(m-2)}{6}z^n y^{m-3} \operatorname{Tr}(\rho[a^+,[a^+,[a^+, H_n]]]) \tag{86}$$

Comparing equations 82 and 86, we easily see that:

$$i(\hat{\dot{g}} - \dot{g}) = \frac{n(n-1)}{2} z^{n-2} y^m \operatorname{Tr}(\rho[a,[a,H_n]]) + \frac{n(n-1)(n-2)}{6} z^{n-3} y^m \operatorname{Tr}(\rho[a,[a,[a,H_n]]])$$
$$- \frac{m(m-1)}{2} z^n y^{m-2} \operatorname{Tr}(\rho[a^+,[a^+,H_n]]) + \frac{m(m-1)(m-2)}{6} z^n y^{m-3} \operatorname{Tr}(\rho[a^+,[a^+,[a^+,H_n]]])$$

(87)

To evaluate this further, we may use two relations proved for any analytic function f in IJBC-45 and 46:

$$[a, f_n(a^+, a)] = \left(\frac{\partial f}{\partial y}\right)_n \tag{88}$$

$$[a^+, f_n(a^+, a)] = -\left(\frac{\partial f}{\partial z}\right)_n \tag{89}$$

Applying these to equation 87, we arrive at the following for any pure state ρ:

$$i(\hat{\dot{g}} - \dot{g}) = \frac{n(n-1)}{2} z^{n-2} y^m \left(\frac{\partial^2 H}{\partial y^2}\right) + \frac{n(n-1)(n-2)}{6} z^{n-3} y^m \left(\frac{\partial^3 H}{\partial y^3}\right)$$
$$- \frac{m(m-1)}{2} z^n y^{m-2} \left(\frac{\partial^2 H}{\partial z^2}\right) - \frac{m(m-1)(m-2)}{6} z^n y^{m-3} \left(\frac{\partial^3 H}{\partial z^3}\right) \tag{90}$$

Finally, for the special case of g in equation 77, we may calculate derivatives of g and use them to simplify equation 90 still further:

$$i(\hat{\dot{g}} - \dot{g}) = \tfrac{1}{2} \frac{\partial^2 g}{\partial z^2} \cdot \frac{\partial^2 H}{\partial y^2} + \tfrac{1}{6} \frac{\partial^3 g}{\partial z^3} \cdot \frac{\partial^3 H}{\partial y^3} - \tfrac{1}{2} \frac{\partial^2 g}{\partial y^2} \cdot \frac{\partial^2 H}{\partial z^2} - \tfrac{1}{6} \frac{\partial^3 g}{\partial y^3} \cdot \frac{\partial^3 H}{\partial z^3} \tag{91}$$

### 3.2.3. Calculation of Discrepancy for Any Analytic g(φ,π)

Finally let us calculate the discrepancy for the case of *any* analytic function g(φ,π). As discussed in section 4.2 of [3], any analytic function g(φ,π) may be expressed as a (convergent) sum of terms of the form $C_\alpha g_\alpha$ where $g_\alpha$ is of the form given in equation 77.

Because equation 91 is linear in g, this immediately tells us that equation 91 is also valid for any analytic function $g_\alpha$!

Our main remaining task is to translate the derivatives in equation 91 from the (y,z) coordinate system to the ($\varphi,\pi$) coordinate system of physical interest here. Because the relation between these two coordinate systems is linear, we do not need to use complex or formal methods. We may first solve equations 19 and 20 to get:

$$\varphi = \frac{1}{\sqrt{2}}(z+y) \tag{92}$$

$$\pi = \frac{1}{i\sqrt{2}}(z-y) \tag{93}$$

Thus any function $g(\varphi,\pi)$ can be represented as a function of z and y:

$$g(\varphi,\pi) = g(\frac{z+y}{\sqrt{2}}, \frac{z-y}{i\sqrt{2}}) \tag{94}$$

Using the chain rule to differentiate with respect to z and y, we get:

$$\frac{\partial g}{\partial z} = \frac{\partial g}{\partial \varphi} \cdot \frac{1}{\sqrt{2}} + \frac{\partial g}{\partial \pi} \cdot \frac{1}{i\sqrt{2}} \tag{95}$$

$$\frac{\partial g}{\partial y} = \frac{\partial g}{\partial \varphi} \cdot \frac{1}{\sqrt{2}} - \frac{\partial g}{\partial \pi} \cdot \frac{1}{i\sqrt{2}} \tag{96}$$

Because of the linear nature of the coordinate transformation here, we may deduce:

$$\frac{\partial}{\partial z} = \frac{\partial}{\partial \varphi} \cdot \frac{1}{\sqrt{2}} + \frac{\partial}{\partial \pi} \cdot \frac{1}{i\sqrt{2}} = \frac{1}{\sqrt{2}}\left(\frac{\partial}{\partial \varphi} - i\frac{\partial}{\partial \pi}\right) \tag{97}$$

$$\frac{\partial}{\partial y} = \frac{\partial}{\partial \varphi} \cdot \frac{1}{\sqrt{2}} - \frac{\partial}{\partial \pi} \cdot \frac{1}{i\sqrt{2}} = \frac{1}{\sqrt{2}}\left(\frac{\partial}{\partial \varphi} + i\frac{\partial}{\partial \pi}\right) \tag{98}$$

From this we may deduce:

$$\frac{\partial^2}{\partial z^2} = \frac{1}{2}\left(\frac{\partial}{\partial \varphi} - i\frac{\partial}{\partial \pi}\right)^2 = \frac{1}{2}\left(\frac{\partial^2}{\partial \varphi^2} - \frac{\partial^2}{\partial \pi^2} - 2i\frac{\partial^2}{\partial \varphi \partial \pi}\right) \tag{99}$$

$$\frac{\partial^2}{\partial y^2} = \frac{1}{2}\left(\frac{\partial}{\partial \varphi} + i\frac{\partial}{\partial \pi}\right)^2 = \frac{1}{2}\left(\frac{\partial^2}{\partial \varphi^2} - \frac{\partial^2}{\partial \pi^2} + 2i\frac{\partial^2}{\partial \varphi \partial \pi}\right) \tag{100}$$

We may use equations 99 and 100 to work out the low-order terms of equation 91 as follows:

$$\begin{aligned}\tfrac{1}{2}\frac{\partial^2 g}{\partial z^2}\cdot\frac{\partial^2 H}{\partial y^2} - \tfrac{1}{2}\frac{\partial^2 g}{\partial y^2}\cdot\frac{\partial^2 H}{\partial z^2} &= \tfrac{1}{8}\left(\frac{\partial^2 g}{\partial \varphi^2} - \frac{\partial^2 g}{\partial \pi^2} - 2i\frac{\partial^2 g}{\partial \varphi \partial \pi}\right)\left(\frac{\partial^2 H}{\partial \varphi^2} - \frac{\partial^2 H}{\partial \pi^2} + 2i\frac{\partial^2 H}{\partial \varphi \partial \pi}\right) \\ &\quad - \tfrac{1}{8}\left(\frac{\partial^2 g}{\partial \varphi^2} - \frac{\partial^2 g}{\partial \pi^2} + 2i\frac{\partial^2 g}{\partial \varphi \partial \pi}\right)\left(\frac{\partial^2 H}{\partial \varphi^2} - \frac{\partial^2 H}{\partial \pi^2} - 2i\frac{\partial^2 H}{\partial \varphi \partial \pi}\right) \\ &= \tfrac{1}{8}\left(-4i\frac{\partial^2 g}{\partial \varphi \partial \pi}\left(\frac{\partial^2 H}{\partial \varphi^2} - \frac{\partial^2 H}{\partial \pi^2}\right) + 4i\left(\frac{\partial^2 g}{\partial \varphi^2} - \frac{\partial^2 g}{\partial \pi^2}\right)\left(\frac{\partial^2 H}{\partial \varphi \partial \pi}\right)\right)\end{aligned}$$

(101)

From equations 97 through 100, we may also deduce:

$$\frac{\partial^3}{\partial y^3} = \frac{1}{\sqrt{2}}\left(\frac{\partial}{\partial \varphi} + i\frac{\partial}{\partial \pi}\right)\frac{1}{2}\left(\frac{\partial^2}{\partial \varphi^2} - \frac{\partial^2}{\partial \pi^2} + 2i\frac{\partial^2}{\partial \varphi \partial \pi}\right) = \frac{1}{2\sqrt{2}}\left(\frac{\partial^3}{\partial \varphi^3} + 3i\frac{\partial^3}{\partial \varphi^2 \partial \pi} - 3\frac{\partial^3}{\partial \varphi \partial \pi^2} - i\frac{\partial^3}{\partial \pi^3}\right) \tag{102}$$

$$\frac{\partial^3}{\partial z^3} = \frac{1}{\sqrt{2}}\left(\frac{\partial}{\partial \varphi} - i\frac{\partial}{\partial \pi}\right)\frac{1}{2}\left(\frac{\partial^2}{\partial \varphi^2} - \frac{\partial^2}{\partial \pi^2} - 2i\frac{\partial^2}{\partial \varphi \partial \pi}\right) = \frac{1}{2\sqrt{2}}\left(\frac{\partial^3}{\partial \varphi^3} - 3i\frac{\partial^3}{\partial \varphi^2 \partial \pi} - 3\frac{\partial^3}{\partial \varphi \partial \pi^2} + i\frac{\partial^3}{\partial \pi^3}\right) \tag{103}$$

Using equations 102 and 103 to work out the higher-order terms in equation 91, we get:

$$\begin{aligned}\tfrac{1}{6}\frac{\partial^3 g}{\partial z^3}\cdot\frac{\partial^3 H}{\partial y^3} - \tfrac{1}{6}\frac{\partial^3 g}{\partial y^3}\cdot\frac{\partial^3 H}{\partial z^3} &= \tfrac{1}{6}\cdot\tfrac{1}{8}\left(\frac{\partial^3 g}{\partial \varphi^3} - 3i\frac{\partial^3 g}{\partial \varphi^2 \partial \pi} - 3\frac{\partial^3 g}{\partial \varphi \partial \pi^2} + i\frac{\partial^3 g}{\partial \pi^3}\right)\left(\frac{\partial^3 H}{\partial \varphi^3} + 3i\frac{\partial^3 H}{\partial \varphi^2 \partial \pi} - 3\frac{\partial^3 H}{\partial \varphi \partial \pi^2} - i\frac{\partial^3 H}{\partial \pi^3}\right) \\ &\quad - \tfrac{1}{6}\cdot\tfrac{1}{8}\left(\frac{\partial^3 g}{\partial \varphi^3} + 3i\frac{\partial^3 g}{\partial \varphi^2 \partial \pi} - 3\frac{\partial^3 g}{\partial \varphi \partial \pi^2} - i\frac{\partial^3 g}{\partial \pi^3}\right)\left(\frac{\partial^3 H}{\partial \varphi^3} - 3i\frac{\partial^3 H}{\partial \varphi^2 \partial \pi} - 3\frac{\partial^3 H}{\partial \varphi \partial \pi^2} + i\frac{\partial^3 H}{\partial \pi^3}\right) \\ &= \tfrac{1}{6}\cdot\tfrac{1}{8}\left(2\left(\frac{\partial^3 g}{\partial \varphi^3} - 3\frac{\partial^3 g}{\partial \varphi \partial \pi^2}\right)\left(3i\frac{\partial^3 H}{\partial \varphi^2 \partial \pi} - i\frac{\partial^3 H}{\partial \pi^3}\right) + 2\left(-3i\frac{\partial^3 g}{\partial \varphi^2 \partial \pi} + i\frac{\partial^3 g}{\partial \pi^3}\right)\left(\frac{\partial^3 H}{\partial \varphi^3} - 3\frac{\partial^3 H}{\partial \varphi \partial \pi^2}\right)\right)\end{aligned}$$

(104)

Based on equation 91, we can calculate $8i(\hat{\dot{g}} - \dot{g})$ by multiplying equations 101 and 104 by 8, and adding them together; the result is identical to equation 67. Q.E.D.

### 3.3 Calculating the Discrepancy for the Case $\varphi, \pi \in \mathbf{R}^2$

The discrepancy can be calculated for the case where $\varphi, \pi \in \mathbf{R}^n$ by using exactly the same procedures as for $\varphi, \pi \in \mathbf{R}^1$, in section 3.2. Unfortunately, the details are complicated and not terribly illuminating. Thus we will present only a few highlights here. Also we will write "a" and "b" instead of "$a_1$" and "$a_2$," in order to reduce the clutter slightly.

To generalize equation 68, we may derive:

$$[a^n b^m, H_n] = [a^n, H_n]b^m + [b^m, H_n]a^n + [a^n, [b^m, H_n]] \tag{105}$$

The first two terms on the right are the same as equation 68 (with "a" and "n" replaced by "b" and "m" for the second term), multiplied on the right by $b^m$ and $a^n$ respectively. The third may be worked out by commuting $a^n$ with the expression from equation 68 for $[b^m, H_n]$. If we assume the extension of equation 66, the last term in equation 105 results in three new cross-terms in the overall expression:

$$mn[a,[b,H_n]]a^{n-1}b^{m-1} + \frac{mn(n-1)}{2}[a,[a,[b,H_n]]]a^{n-2}b^{m-1} + \frac{m(m-1)n}{2}[a,[b,[b,H_n]]]a^{n-1}b^{m-2} \tag{106}$$

Then, following the procedures of section 3.2.2, we arrive at an extension of equation 91:

$$i(\hat{\dot{g}} - \dot{g}) = \tfrac{1}{2}\frac{\partial^2 g}{\partial z_1^2} \cdot \frac{\partial^2 H}{\partial y_1^2} + \tfrac{1}{6}\frac{\partial^3 g}{\partial z_1^3} \cdot \frac{\partial^3 H}{\partial y_1^3} - \tfrac{1}{2}\frac{\partial^2 g}{\partial y_1^2} \cdot \frac{\partial^2 H}{\partial z_1^2} - \tfrac{1}{6}\frac{\partial^3 g}{\partial y_1^3} \cdot \frac{\partial^3 H}{\partial z_1^3}$$

$$+ \tfrac{1}{2}\frac{\partial^2 g}{\partial z_2^2} \cdot \frac{\partial^2 H}{\partial y_2^2} + \tfrac{1}{6}\frac{\partial^3 g}{\partial z_2^3} \cdot \frac{\partial^3 H}{\partial y_2^3} - \tfrac{1}{2}\frac{\partial^2 g}{\partial y_2^2} \cdot \frac{\partial^2 H}{\partial z_2^2} - \tfrac{1}{6}\frac{\partial^3 g}{\partial y_2^3} \cdot \frac{\partial^3 H}{\partial z_2^3}$$

$$+ \frac{\partial^2 g}{\partial z_1 \partial z_2} \cdot \frac{\partial^2 H}{\partial y_1 \partial y_2} - \frac{\partial^2 g}{\partial y_1 \partial y_2} \cdot \frac{\partial^2 H}{\partial z_1 \partial z_2} + \tfrac{1}{2}\frac{\partial^3 g}{\partial z_1^2 \partial z_2} \cdot \frac{\partial^3 H}{\partial y_1^2 \partial y_2} - \tfrac{1}{2}\frac{\partial^3 g}{\partial y_1^2 \partial y_2} \cdot \frac{\partial^3 H}{\partial z_1^2 \partial z_2}$$

$$+ \tfrac{1}{2}\frac{\partial^3 g}{\partial z_1 \partial z_2^2} \cdot \frac{\partial^3 H}{\partial y_1 \partial y_2^2} - \tfrac{1}{2}\frac{\partial^3 g}{\partial y_1 \partial y_2^2} \cdot \frac{\partial^3 H}{\partial z_1 \partial z_2^2}$$

(107)

By inspecting these results and derivations, we can see that the case n>2 would yield exactly the same terms (but using all the additional new index combinations available) plus new terms of the form:

$$\frac{\partial^3 g}{\partial z_1 \partial z_2 \partial z_3} \cdot \frac{\partial^3 H}{\partial y_1 \partial y_2 \partial y_3} - \frac{\partial^3 g}{\partial y_1 \partial y_2 \partial y_3} \cdot \frac{\partial^3 H}{\partial z_1 \partial z_2 \partial z_3} \qquad (108)$$

The new cross-terms in equations 107 and 108 are all third-order terms, except for the following term which we can work out by using the methods of section 3.2.3:

$$\frac{\partial^2 g}{\partial z_1 \partial z_2} \cdot \frac{\partial^2 H}{\partial y_1 \partial y_2} - \frac{\partial^2 g}{\partial y_1 \partial y_2} \cdot \frac{\partial^2 H}{\partial z_1 \partial z_2} = -2i\left(\frac{\partial^2 g}{\partial \varphi_1 \partial \pi_2} + \frac{\partial^2 g}{\partial \varphi_2 \partial \pi_1}\right)\left(\frac{\partial^2 H}{\partial \varphi_1 \partial \varphi_2} - \frac{\partial^2 H}{\partial \pi_2 \partial \pi_1}\right)$$
$$+ 2i\left(\frac{\partial^2 g}{\partial \varphi_1 \partial \varphi_2} - \frac{\partial^2 g}{\partial \pi_2 \partial \pi_1}\right)\left(\frac{\partial^2 H}{\partial \varphi_1 \partial \pi_2} + \frac{\partial^2 H}{\partial \varphi_2 \partial \pi_1}\right) \qquad (109)$$

## 4. When Does the Discrepancy (67) Equal Zero in Equilibrium States?

Section 1 has already discussed the larger implications of equation 67. There are many possible neoclassical models to try to explain quantum dynamics. This paper is exploring only one particular version, in which the Heisenberg operators are assumed to operate on the classical density matrix as defined in equation 7, *without* projection (as in equations 1 and 2), and in which observables g($\varphi$, $\pi$) are identified with the *normal form* field operator $g_n(\Phi, \Pi)$, not the usual $g(\Phi, \Pi)$.

Equation 67 yields the discrepancy in predictions of the flux in observables between QFT and this neoclassical theory in the general case for pure states $\rho(\varphi,\pi)$. Because of the linearity in $g_n$, we can calculate the expected value of the discrepancy for a statistical ensemble ($\rho \in PR$) simply by taking the expected value of equation 67.

For equivalence between QFT and this particular neoclassical theory, as discussed in section 1, we would ask for any IEE that:

$$<\hat{\dot{g}} - \dot{g}> = <\hat{\dot{g}}> = 0 \tag{110}$$

The requirement that $\partial_t^n <g> = 0$ in any neoclassical IEE provides a substantial amount of information which might be used in an effort to prove that a variety of expressions similar to the right-hand side of equation 67 must be zero. (In the PDE case, definiteness of momentum and translational invariance of the distribution are also important.) However, it is easy to see that this by itself would not be enough to prove equation 110 in the general case.

Consider the following example. Pick the observable:

$$g(\varphi,\pi) = \varphi\pi \tag{111}$$

and pick the classical harmonic oscillator Hamiltonian:

$$H = \tfrac{2}{2}\pi^2 + \tfrac{1}{2}m\varphi^2 \tag{112}$$

In this case, the discrepancy shown in equation 67 reduces to:

$$<8i(\hat{\dot{g}} - \dot{g})> = -4i(m-1) \tag{113}$$

Clearly there would be no way to deduce that this must equal zero in the general case for any matrix ρ which represents an IEE! In fact, the condition cannot be met at all when m≠1!

On the other hand, this example is not terribly disturbing. QFT goes to great pains to make sure that its starting points are always formulated *in such a way* that $\Phi_j$ and $\Pi_j$ are canonically conjugate. In effect, the theory is scaled, using terms like sqrt(m), to achieve a standard, normalized balance between $\varphi_j$ and $\pi_j$. Given any harmonic oscillator system, as in equation 112, we can represent the dynamics equivalently by performing a *field scaling*, which results in m=1; after that, we arrive at zero discrepancy in equation 67 for the scaled field, *for all states* ρ, not just equilibria! The field scaling here is essentially just a form of mass renormalization.

For more general Hamiltonians, we cannot make the corresponding term in equation 67 equal zero *for all states* ρ. However, we *can* scale the field so that:

$$\left\langle \frac{\partial^2 H}{\partial \varphi^2} - \frac{\partial^2 H}{\partial \pi^2} \right\rangle = 0 \qquad (114)$$

for an "elementary particle" or "basic soliton" kind of state of *given mass energy*. This may sound like cheating at first – but consider that QFT itself must be renormalized according to the *actual physical mass* of *each* elementary particle appearing in the theory in order to work! In the PDE case, this can be applied to each of the input and output scattering channels, in order to establish equivalence.

At first, this extension to the PDE case may seem to entail an additional problem: what about the fact that a given type of elementary particle still has a different mass-energy depending on its velocity? Here again, we may look to QFT for the answer.

In the PDE case (discussed in [3] more concretely), QFT does not really start from the corresponding CFT as such. Instead, it starts from integro-differential equations resulting after scaling according to sqrt(w) after a Fourier analysis, where w represents the *actual* mass-energy of the particle.

These kinds of considerations should be enough to address the first term in equation 67, roughly speaking. What about the other terms? The second term refers to a kind of term which does not exist in Hamiltonians used in QFT (because it would drop out in the dynamics anyway); likewise for the corresponding term in equation 109. The third-order terms in equation 67 all refer to the impact on the predictions of QFT of $[a,[a,[a,H_n]]]$ or $[a^+,[a^+,[a^+,H_n]]]$; however, the impact of these terms is to insert triple-creation-from-vacuum and triple-destruction-to-vacuum terms in Feynman diagrams! In actual calculations in QED[16,17], such terms are all simply thrown out. Their effect on physics has not really been empirically validated, so far as we know.

Expectation values of equation 67 for higher-order functions g presumably would not be zero without *additional* field scaling. Could a few simple types of field scaling take care of all the higher-order terms involved in these expansions? It would be reasonable to expect that the answer here would depend on the form of the Hamiltonian $H_n$. In fact, there is an obvious parallel here to the process of renormalization, in which an entire Taylor series can be accommodated, *so long as* the relevant Hamiltonian is finite or renormalizable, and so long as we make use of several basic forms of renormalization or field scaling which can also be used here – mass scaling (as just discussed), charge renormalization or scaling, and form factors (integro-differential equation representation).

It seems quite plausible, then, that the discrepancies will actually go to zero, for the case of properly renormalized bosonic field theories. Even if they do not go to zero, the discrepancies may well be small enough to justify consideration of this neoclassical theory as an alternative to traditional quantum dynamics.

**Appendix. Reification of Statistics, An Alternative Approach**

The statistics of classical field theories have been studied for many decades. In the past, it was difficult to compare these statistics with quantum dynamics, because of a well-known problem often called "the closure of turbulence." This is closely related to the "problem of moments" and the problem of unrealistic matrices ρ discussed in section 1. This paper mainly tries to evade these problems, by focusing on the dynamics of observables g rather than the dynamics of the moments themselves. The free space master equations given in equations 12, 13 and 58 still suffer in principle from the problem of "closure of turbulence."

In two previous papers [5,20], we have suggested an approach called "reification" to solve the problem of closure here. More precisely, we described a way to recode the moments, so as to maintain the essential unitary character of the dynamics for ρ∈PR but to eliminate the appearance of non-unitary dynamics due to unrealistic states. This results in new dynamics which *can* be represented in terms of a wave equation or a Liouville equation, based on an Hermitian "Hamiltonian" matrix – but the new Hamiltonian matrix need not be the same as $H_n(\underline{\Phi}, \underline{\Pi})$ for the original classical field! We end up with *zero* discrepancies between the classical field theory and the operator field theory, but we are not guaranteed that the resulting operator field theory "corresponds to" the same classical field theory as the one we started with!

Reification *does not appear to work* for the classical density matrix ρ as defined in equation 7. However, the difficulties do appear to go away when we use an expanded version of the classical density matrix, as described in the Appendix of IJBC[3]. The rest of this Appendix will give the details.

To begin with, consider the case with ρ defined as in equation 7, for the case of φ, π ∈ $R^1$. Let us define a Hermitian reification operator:

$$S(\alpha) = \exp(-\frac{\alpha}{2}(a^+a^+ + aa)) \tag{A.1}$$

where α is real. S induces a similarity transformation. For example:

$$S(d\alpha)aS^{-1}(d\alpha) = (1-\frac{d\alpha}{2}(a^+a^+ + aa))a(1+\frac{d\alpha}{2}(a^+a^+ + aa)) + o(d\alpha)$$
$$= 1 + a^+ d\alpha + o(d\alpha) \tag{A.2}$$

Likewise:

$$S(d\alpha)a^+ S^{-1}(d\alpha) = 1 - ad\alpha + o(d\alpha) \tag{A.3}$$

In short, S induces a kind of rotation between creation and annihilation operators.

If we chose a rotation of α=(π/4), we might expect to get:

$$a \to \frac{1}{\sqrt{2}}(a + a^+) \tag{A.4}$$

$$a^+ \to \frac{1}{\sqrt{2}}(a^+ - a) \tag{A.5}$$

We could then define a transformed density matrix:

$$\rho_z = S(\tfrac{\pi}{4})\rho S(\tfrac{\pi}{4}) \tag{A.6}$$

But in that case we could derive:

$$\rho_z\left(\frac{a+a^+}{\sqrt{2}}\right) = (S\rho S)(S^{-1}a^+ S) = (\varphi - i\pi)\rho \tag{A.7}$$

$$\left(\frac{a+a^+}{\sqrt{2}}\right)\rho_z = (SaS^{-1})(S\rho S) = (\varphi + i\pi)\rho \tag{A.8}$$

Clearly A.7 and A.8 cannot both be true in the general case!

The obvious explanation here is that S(π/4) results in an unbounded or even undefined matrix $\rho_z$. Unbounded matrices over Fock space lead to calculations dependent on infinite series which do not possess absolute convergence properties, and hence are ill-defined. (Note, by contrast, that the classical density matrix defined here is very well-behaved, so long as the original vectors **φ** and **π** possess a finite norm.) In other words, the usual theorems of operator theory used in QFT no longer apply, because the conditions on those theorems are violated.

One may verify that this is a plausible explanation, very roughly, by first defining:

$$\rho_z(\alpha) = S(\alpha)\rho S(\alpha) \tag{A.9}$$

$$\rho(\alpha) = \frac{\rho_z(\alpha)}{\text{Tr}(\rho_z(\alpha))} \tag{A.10}$$

We know that $\rho(\alpha)$ must obey a differential equation in $\alpha$ the same as what A.9 implies for $\rho_z$, except for terms which multiply it by a scalar. In order to keep the norm of $\rho(\alpha)$ constant as we change $\alpha$, we basically need to pick the parameter functions $c(\alpha)$ and $d(\alpha)$ in:

$$\frac{\partial \rho}{\partial \alpha} = -\frac{a^+ a^+ + aa}{2}\rho - \rho\frac{a^+ a^+ + aa}{2} + c(\alpha)a^2(\alpha)\rho + c(\alpha)\rho(a^H(\alpha))^2 + d(\alpha)a(\alpha)\rho a^H(\alpha) \tag{A.11}$$

where:

$$a(\alpha) = (\cos\alpha)a + (\sin\alpha)a^+ \tag{A.12}$$

Working out the algebra of what is needed to keep $\text{Tr}(\rho)$ constant, we arrive roughly at:

$$c(\alpha) = \frac{1}{1 - 4\sin^2\alpha\cos^2\alpha} \tag{A.13}$$

$$d(\alpha) = \frac{-4\sin\alpha\cos\alpha}{1 - 4\sin^2\alpha\cos^2\alpha} \tag{A.14}$$

We have not been so careful with details here as in the body of this paper, since a rough calculation is enough for present purposes: we see very clearly that as $\alpha$ approaches $\pi/4$, the equation blows up, suggesting that the original $\rho_z(\alpha)$ does indeed become unbounded or even ill-defined as $\alpha \to \pi/4$. It is interesting that it blows up precisely at those values of $\alpha$ which would have made the dynamics unitary! This may possibly be related to the fact that the dynamics *cannot be made* unitary because the free space master equation may contain eigenvalues $\lambda$ with $\text{Re}(\lambda) \neq 0$, in the general case, for matrices $\rho \notin \text{PR}$. It may also be related to the fact that we can often find equilibrium ensembles $\rho$ arbitrarily close to each other, with slightly different energies, which would have to be made orthogonal to each other after any such reification.

On the other hand, consider what happens when we define an expanded version of ρ based on replacing equation 6 by:

$$\underline{\tilde{w}}(\underline{\varphi},\underline{\pi}) \approx \exp\left(\frac{1}{\sqrt{2}} \sum_{j=1}^{n} \left((\varphi_j + i\pi_j)a_j^+ + (\varphi_j - i\pi_j)b_j^+ - (\varphi_j^2 + \pi_j^2)\right)\right)|0\rangle \quad (A.15)$$

as suggested in the Appendix of IJBC [3]. (We have put an approximate-equals sign here because we have not performed enough calculations with this definition to know the cleanest way to insert factors of sqrt(2).) Consider what happens when we perform reification based on:

$$M(\alpha) = \exp\left(-\alpha \sum_{j=1}^{n}(a_j^+ b_j^+ + a_j b_j)\right) \quad (A.16)$$

Reification based on M(α) should be valid if it does not become unbounded for matrices ρ ∈ PR. We have not yet proved this; however, in working through the same algebra as with S, it now appears that the paradox has disappeared! In this expanded representation, there are many ways to write dynamical equations for ρ which are mutually equivalent and valid for ρ ∈ PR, but different both in appearance and in their impact on ρ ∉ PR. For example, if we define:

$$\underline{z}(\underline{\varphi},\underline{\pi}) = M(\tfrac{\pi}{4})\underline{w}(\underline{\varphi},\underline{\pi}) \quad (A.17)$$

we may derive a wave equation in the general case of the form:

$$\underline{\dot{z}} = \left(\sum Af(\underline{\Phi},\underline{\Pi})\right)\underline{z} \quad (A.18)$$

where the sum is over different antiHermitian operators A and analytic functions f of (here) commuting Hermitian operators $\underline{\Phi}$ and $\underline{\Pi}$ (not the same as the operators of that name in the body of this paper!). In many situations (e.g. [20]) the overall terms in parentheses will be antiHermitian, but we have yet to analyze the general case in detail. This is presented as a *possible* alternative to the approach given in the body of this paper, but we do not yet know how useful or realistic it may be.